# Historical Note on Relativistic Theories of Electromagnetism

Valeri V. Dvoeglazov
Escuela de Fisica, Universidad Autonoma de Zacatecas
Apartado Postal C-580, Zacatecas 98068, ZAC., Mexico
Internet address: valeri@cantera.reduaz.mx
URL: http://cantera.reduaz.mx/~ valeri/valeri.htm

Quantum electrodynamics is the well-accepted theory. However, we feel it is useful to look at formalisms that provide alternative ways to describe light, because in the recent years the development of quantum field theories based primarily on the gauge principle has encountered considerable difficulties. There is a wide variety of generalized theories, and they are characterized mainly by the introduction of additional parameters and/or longitudinal modes of electromagnetism. The Majorana-Oppenheimer form of electrodynamics, the Sachs theory of Elementary Matter, the analysis of the action-at-a-distance concept, presented recently by Chubykalo and Smirnov-Rueda, and the analysis of the claimed 'longitudinality' of the antisymmetric tensor field after quantization are reviewed in this essay. We also list recent advances in the Weinberg $2(2J + 1)$ formalism (which is built on First Principles) and in the Majorana theory of neutral particles. These may serve as starting points for constructing a quantum theory of light.

Maxwell's electromagnetic theory perfectly describes many observed phenomena. The accuracy of the predictions of quantum electrodynamics is without precedent [1]. These are widely accepted as the only tools for dealing with electromagnetic phenomena. Other modern field theories have been built on the basis of similar principles to treat weak, strong and gravitational interactions. Nevertheless, many scientists have felt a certain dissatisfaction with both of these theories, almost since their inception, see, *e.g.*, refs. [2] and refs. [4-6]. In the preface to the Dover edition of his book [7] A. Barut writes (1979): "Electrodynamics and the classical theory of fields remain very much alive and continue to be the source of inspiration for much of the modern research work in new physical theories." And in the preface to the first edition he speaks of the shortcomings of modern quantum field theory. These are well known. Furthermore, in spite of great expectations in the sixties and seventies after the proposal of the Glashow-Salam-Weinberg model and quantum chromodynamics, attempts to formulate a unified field theory based on the gauge principle have run into serious difficulties.

At the end of the nineties, we now have considerable experimental data at our disposal which are not satisfactorily explained on the basis of the standard model. First of all, we may single out the following: the LANL neutrino oscillation experiment; the atmospheric neutrino anomaly, the solar neutrino puzzle (all of the above-mentioned imply existence of the neutrino mass); the tensor coupling in decays of $p^-$ and $K^+$ mesons; the dark matter problem; the observed periodicity of the number distribution of galaxies, and the 'spin crisis' in QCD. Added to this are experiments and observations involving superluminal phenomena: negative mass-square neutrinos, tunnelling photons, X-shaped waves and superluminal expansion in quasars and galactic objects.

In the meantime, since the time the Lorentz-Poincaré-Einstein Theory of Relativity [8] was proposed and the mathematical formalism of the Poincaré group [9] was introduced, many physicists (including A. Einstein, W. Pauli and M. Sachs) have felt that in order to build a reliable theory (which would be based on relativistic ideas) one must utilize the irreducible representations of the underlying symmetry group—the Poincaré group of special relativity—and the Principle of Causality, *i.e.* it must be built from first principles. Considerable effort has been made recently in this direction [10-17]. Since the prediction and discovery of an additional phase-free variable, spin, which all observed fundamental particles have, finding its classical analogue and relating it to known fields and/or space-time structures (perhaps in higher dimensions) has been one of the chief tasks of physicists. Understanding the nature of mass, the parity violation effect on the kinematical level and the reasons for the different scales of different interactions has been on the list as well. We can now say that some progress has been achieved. At the end of this introductory part we note that although the Ultimate Theory has not yet been proposed, the recent papers of D. V. Ahluwalia, M. W. Evans[*], E. Recami and several other works provide a sufficiently clear way to this goal.

We deal first with the historical development and ideas that may prove useful in making further progress.

***E* = 0 *solutions.*** First of all, I would like to mention the problem of existence of 'acausal' solutions of relativistic wave equations of the first order. In ref. [10] and then in [11] the author, D. V. Ahluwalia, found that massless equations of the form[†]

$$(\mathbf{J} \cdot \mathbf{p} - p_o 1) f_R(\mathbf{p}) = 0, \tag{1a}$$

$$(\mathbf{J} \cdot \mathbf{p} + p_o 1) f_L(\mathbf{p}) = 0 \tag{1b}$$

have acausal dispersion relations, see Table 2 in [10]. In the case of the spin $j = 1$ this manifests in existence of the solution with the energy $E = 0$. Some time ago we learned that the same problem has been discussed by J. R. Oppenheimer [18], S. Weinberg [19b] and E. Gianetto [20c]. For instance, Weinberg has indicated that

> "for $j = \frac{1}{2}$ [the equations (1a,1b)] are the Weyl equations for the left- and right-handed neutrino fields, while for $j = 1$ they are just Maxwell's free-space equations for left- and right- circularly polarized radiation:
>
> $$\nabla \times [\mathbf{E} - i\mathbf{B}] + i \frac{\partial}{\partial t}[\mathbf{E} - i\mathbf{B}] = 0, \tag{2a}$$

---

[*] Although I frequently disagree with Dr. M.W. Evans, his main idea is reasonable.

[†] Here and below in this historical essay we try to keep the notation and the metric of original papers.

$$\nabla \times [\mathbf{E} + i\mathbf{B}] - i\frac{\partial}{\partial t}[\mathbf{E} + i\mathbf{B}] = 0 \tag{2b}$$

*The fact that these field equations are of first order for any spin seems to me to be of no great significance, since in the case of massive particles we can get along perfectly well with (2j+1)–component fields which satisfy only the Klein-Gordon equation."*

This is obviously a remarkable and bold conclusion by this great physicist.

Oppenheimer was also concerned with the $E = 0$ solution [p.729,730,733,735] [18] and he indicated at its connection with the electrostatic solutions of Maxwell's equations. "In the absence of charges there may be no such field." At first sight this seems contradictory: free-space Maxwell's equations do not contain $r_e$ or $r_m$ terms, the charge densities, but dispersion relations still tell us about the solution $E = 0$. He deals further with the matters of relativistic invariance of the matrix equation (p. 733) and suggests that the components of $\psi$ ($f_{R,L}$ in the notation of [10,11]) transform under pure Lorentz transformations like the space components of a covariant 4-vector. This induces him to extend the matrices and the wave functions to include the fourth component. A similar formulation was developed by Majorana [20]. If so, it would be already difficult to consider $f_{R,L}$ as Helmoltz bi-vectors because they have different laws for pure Lorentz transformations. What does the 4-component function (and its space components) correspond to? Finally, he indicated (p. 728) that $c\mathbf{t}$, the angular momentum matrices, and the corresponding density-flux vector may "play in some respects the part of the velocity", with eigenvalues 0, $\pm c$. Thus, in my opinion, the formula (5) of the paper [18] may have some relations with the discussion of the convection displacement current in [3], see below.

Finally, M. Moshinsky and A. Del Sol found a solution of similar nature in a two-body relativistic problem [21]. Of course, it is connected with earlier considerations, *e.g.*, with the problem of the relative time in the quasipotential approach. In order to try to understand the physical sense of the $E = 0$ solutions and the corresponding field components, let us consider other generalizations of the Maxwell formalism.

*The 'baroque' formalism.* In this formalism, proposed in the fifties by K. Imaeda [22] and T. Ohmura [23], who intended to solve the problem of electron stability, additional scalar and pseudoscalar fields are introduced in Maxwell's theory. Monopoles and magnetic currents are also present in this theory. The equations become the following:

$$\mathrm{rot}\,\mathbf{H} - \frac{\partial \mathbf{E}}{\partial x_o} = \mathbf{i} - \mathrm{grad}\,e, \tag{3a}$$

$$\mathrm{rot}\,\mathbf{E} + \frac{\partial \mathbf{H}}{\partial x_o} = \mathbf{i} + \mathrm{grad}\,h, \tag{3b}$$

$$\mathrm{div}\,\mathbf{E} = r + \frac{\partial e}{\partial x_o}, \tag{3c}$$

$$\mathrm{div}\,\mathbf{H} = -s + \frac{\partial h}{\partial x_o}. \tag{3d}$$

"Each of **E** and **H** is separated into two parts $\mathbf{E}^{(1)} + \mathbf{E}^{(2)}$ and $\mathbf{H}^{(1)} + \mathbf{H}^{(2)}$: one is the solution of the equations with **j**,**s**, *h* zero, and other is the solution of the equations with **i**, **r**, *e* zero." Furthermore, T. Ohmura indicated the existence of longitudinal photons in her model: "It will be interesting to test experimentally whether the *g*ray keeps on its transverse property even in the high energy region as derived from the Maxwell theory or it does not as predicted from our hypothesis." In fact, equations (3a)-(3d) can be written in matrix notation, which leads to the known Majorana-Oppenheimer formalism for the $(0,0) \oplus (1,0)$ [or $(0,0) \oplus (0,1)$] representation of the Poincaré group [20,18], see also [24]. In a form with the Majorana-Oppenheimer matrices

$$r^1 = \begin{pmatrix} 0 & -1 & 0 & 0 \\ -1 & 0 & 0 & 0 \\ 0 & 0 & 0 & -i \\ 0 & 0 & i & 0 \end{pmatrix}, \quad r^2 = \begin{pmatrix} 0 & 0 & -1 & 0 \\ 0 & 0 & 0 & i \\ -1 & 0 & 0 & 0 \\ 0 & -i & 0 & 0 \end{pmatrix}, \tag{4a}$$

$$r^3 = \begin{pmatrix} 0 & 0 & 0 & -1 \\ 0 & 0 & -i & 0 \\ 0 & i & 0 & 0 \\ -1 & 0 & 0 & 0 \end{pmatrix}, \quad r^o = 1_{4\times 4}, \tag{4b}$$

and $\bar{r}^o \equiv r^o$, $\bar{r}^i \equiv -r^i$ the equations without an explicit mass term are written

$$(r^m \P_m) y_1(x) = f_1(x), \tag{5a}$$

$$(\bar{r}^m \P_m) y_2(x) = f_2(x). \tag{5b}$$

The $f_i$ are the "quadrivectors" of the sources

$$f_1 = \begin{pmatrix} -r + i\mathbf{s} \\ i\mathbf{j} - \mathbf{i} \end{pmatrix}, \quad f_2 = \begin{pmatrix} r + i\mathbf{s} \\ -i\mathbf{j} - \mathbf{i} \end{pmatrix} \tag{6}$$

The field functions were considered to be

$$y_1(p^m) = C y_2^*(p^m) = \begin{pmatrix} -i(E^o + iB^o) \\ E^1 + iB^1 \\ E^2 + iB^2 \\ E^3 + iB^3 \end{pmatrix}, \quad y_2(p^m) = C y_1^*(p^m) = \begin{pmatrix} -i(E^o + iB^o) \\ E^1 - iB^1 \\ E^2 - iB^2 \\ E^3 - iB^3 \end{pmatrix}, \tag{7}$$

where $E^o \equiv -h$, $B^o \equiv e$ and

$$C = C^{-1} = \begin{pmatrix} -1 & 0 & 0 & 0 \\ 0 & 1 & 0 & 0 \\ 0 & 0 & 1 & 0 \\ 0 & 0 & 0 & 1 \end{pmatrix}, \quad C a^m C^{-1} = \bar{a}^{m*}. \tag{8}$$

When sources are switched off the equations have relativistic dispersion relations $E = \pm |\mathbf{p}|$ only. In ref. [20] zero-components of $y$ have been connected with $p_o = \P_m A^m$, the zero-component of the canonically conjugate momentum to the field $A_m$. H. E. Moses developed the Oppenheimer's idea [18] that the longitudinal part of the electromagnetic field is connected somehow with the sources which created it [Eq.(5.21)] [25]. Moreover, it was mentioned in this work that even after the switchoff of the sources, the localized field can possess the longitudinal component (*Example 2*). Then, he made a convention which, in my opinion, is required to give more rigorous scientific basis: "... $y^A$ is not suitable for a final field because it is not purely transverse. Hence we shall subtract the part whose divergence is not zero."

Finally, we should mention ref. [26]. The proposed formalism is connected with the formalism of the previously cited works (and with the massive Proca theory). Two of Maxwell's equations remain unchanged, but one has additional terms in two other ones:

$$\nabla \times \mathbf{H} - \frac{\partial \mathbf{D}}{\partial t} = \mathbf{J} - \frac{1}{m_o \ell^2} \mathbf{A}, \tag{9a}$$

$$\nabla \cdot \mathbf{D} = r - \frac{e_o}{\ell^2} V, \tag{9b}$$

where $\ell$ is of the dimensions length and is suggested by Lyttleton and Bondi to be of the order of the radius of the Universe. $\mathbf{A}$ and $V$ are the vector and scalar potentials, which put back into two Maxwell's equations for strengths. So, these additional terms contain information about possible effects of the photon mass. This was applied to explain the expansion of the Universe. The Watson generalization, also discussed in [26b], is based on the introduction of the additional gradient current [as in Eqs. (3a, 3c)] and, in fact, repeats in essence the Majorana-Oppenheimer and Imaeda-Ohmura formulations. On a scale much smaller than a radius of the Universe, both formulations were shown by Chambers to be equivalent. The difference obtained is of order $\ell^{-2}$ at the most. In fact, both formulations were noted by Chambers to be able to describe local creation of the charge. The question of the integral conservation of the charge over the volume still deserves elaboration, the question of possibility to observe such a type of non-conservation as well. These questions may be connected with the boundary conditions on the sphere of the radius $\ell$.

*The theory of Elementary Matter.* The formalism proposed by M. Sachs [27,28] is on the basis of the consideration of "spinorial" functions composed of 3-vector components:

$$\mathbf{f}_1 = \begin{pmatrix} G_3 \\ G_1 + iG_2 \end{pmatrix}, \quad \mathbf{f}_2 \begin{pmatrix} G_1 - iG_2 \\ -G_3 \end{pmatrix}, \tag{10}$$

where $G_k = H_k + iE_k$ ($k = 1,2,3$). 2-component functions of the currents are constructed in the following way:

$$\Upsilon_1 = -4p\,i \begin{pmatrix} r + j_3 \\ j_1 + ij_2 \end{pmatrix}, \quad \Upsilon_2 = -4p\,i \begin{pmatrix} j_1 - ij_2 \\ r - j_3 \end{pmatrix}. \tag{11}$$

The dynamical equation in this formalism reads

$$\boldsymbol{s}^m \partial_m \boldsymbol{f}_a = \Upsilon_a. \tag{12}$$

> *... Eq. (12) is* not *equivalent to the less general form of Maxwell's equations. That is to say the spinor equations (12) are not merely a rewriting of the vector form of the field equations, they are a true generalization in the sense of transcending the predictions of the older form while also agreeing with all of the correct predictions of the* *latter ... Eq. (12) may be rewritten in the form of four conservation equations* $\partial_m\left(f_a^\dagger s^m f_b\right) = f_a^\dagger \Upsilon_b + \Upsilon_a^\dagger f_b$ *[which] entails eight real conservation laws.*

For instance, these equations could serve as a basis for describing parity-violating interactions [27a], and can account for the spin-spin interaction as well from the beginning [27d,p.934]. The formalism was applied to explain several puzzles in neutrino physics. The connection with the Pauli Exclusion Principle was revealed. The theory, when the interaction ('matter field labeling') is included, is essentially bi-local.[‡]

> *"What was discovered in this research program, applied to the particle-antiparticle pair, was that an exact solution for the coupled field equations for the pair, in its rest frame, gives rise (from Noether's theorem) to a prediction of null energy, momentum and angular momentum, when it is in this particular* bound state." *[28]*

Later [28] this type of equations was written in the quaternion form with the continuous function $m = l\hbar/c$ identified with the inertial mass. Thus, an extension of the model to the general relativity case was proposed. Physical consequences of the theory are: a) the formalism predicts while small but non-zero masses and the infinite spectrum of neutrinos; b) the Planck spectral distribution of black body radiation follows; c) the hydrogen spectrum (including the Lamb shift) was deduced; d) bases for the charge quantization are proposed; e) the lifetime of the muon state was predicted; f) the electron-muon mass splitting was discussed,

> *"... the difference in the mass eigenvalues of a doublet depends on the alteration of the geometry of space-time in the vicinity of excited pairs of the 'physical vacuum' ['a degenerate gas of spin-zero objects,' longitudinal and scalar photons, in fact!* V. V. D.*]—leading, in turn, to a dependence of the ratio of mass eigenvalues on the fine-structure constant".*

That was impressive work and these are impressive results!

*Quantum mechanics of phase.* A. Staruszkiewicz [29,30] considers the Lagrangian and the action of a potential formulation for the electromagnetic field, which include a longitudinal part:

$$S = -\frac{1}{16\pi} \int d^4x \left\{ F_{mn} F^{mn} + 2g \left( \partial^m A_m + \frac{1}{e} \partial_m \partial^m S \right)^2 \right\}. \tag{13}$$

*S* is a scalar field called the phase. As a matter of fact, this formulation was shown to be a development of the Dirac-Fock-Podol'sky model in which the current is a gradient of some scalar field [31]:

$$4\pi j_n = -\partial_n F. \tag{14}$$

The modified Maxwell's equations are written:

$$\partial_l F_{mn} + \partial_m F_{nl} + \partial_n F_{lm} = 0, \tag{15a}$$

---

[‡] The hypothesis of the non-local nature of charge seems to have been first proposed by J. Frenkel.

$$\partial^m F_{mn} + \partial_n F = 0 \,. \tag{15b}$$

Again we see a gradient current and, therefore, the Dirac-Fock-Podol'sky model is a simplified version (apparently without monopoles) of the more general Majorana-Oppenheimer theory. Staruszkiewicz posed the following questions [30], see also [23b] and [32]: "Is it possible to have a system, whose motion is determined completely by the charge conservation law alone? Is it possible to have a pure charge not attached to a nonelectromagnetic piece of matter?" and answering came to the conclusion "that the Maxwell electrodynamics of a gradient current is a closed dynamical system." The interpretation of a scalar field as a phase of the expansion motion of a charge under repulsive electromagnetic forces was proposed. "They [the Dirac-Fock-Podol'sky equations] describe a charge let loose by removal of the Poincaré stresses." The phase was then related with the vector potential by means of [30e,p.902]

$$S(x) = -e \int A_m(x-y) j^m(y) \mathrm{d}^4 y, \quad \partial_m j^m(y) = \partial^{(4)}(y) \,. \tag{16}$$

Formula (16) is reminiscent to the Barut self-field electrodynamics [33]. This should be investigated by taking the 4-divergence of Barut's *anzatz*.

Next, the operator of a number of zero-frequency photons was studied. The total charge of the system, found on the basis of the Noether theorem, was connected with the change of the phase between the positive and the negative time-like infinity: $Q = -\frac{e}{4p}[S(+\infty) - S(-\infty)]$. It was shown that $e^{iS}$, having a Bose-Einstein statistics, can serve itself as a creation operator: $Q e^{iS}|0\rangle = [Q, e^{iS}]|0\rangle = -e\, e^{iS}|0\rangle$. Questions of fixing the factor $g$ by appropriate physical conditions were also answered. Finally, the Coulomb field was decomposed into irreducible unitary representations of the proper orthochronous Lorentz group [34]. Both representations of the main series and the supplementary series were regarded. In my opinion, this research can help to understand the nature of the charge and of the fine structure constant.

*Invariant evolution parameter.* The theory of electromagnetic field with an invariant evolution parameter ($t$, the Newtonian time) has been worked out by L. P. Horwitz [35-37]. It is a development of the Stueckelberg formalism [38] and I consider this theory as an important step to understanding the nature of our space-time. The Stueckelberg equation

$$i \frac{\partial y_t(x)}{\partial t} = K y_t(x) \tag{17}$$

is deduced on the basis of his worldline classical relativistic mechanics with following setting up the covariant commutation relations $[x^m, p^n] = i g^{mn}$. Remarkably, he proposed a classical analogue of antiparticle (which, in fact, has been later used by R. Feynman) and of annihilation processes. As noted by Horwitz if one insists on the $U(1)$ gauge invariance of the theory based on the Stueckelberg-Schrödinger equation (17) one arrives at the 5-potential electrodynamics ($i\partial_t \to i\partial_t + e_o a_5$) where the equations, which are deduced by means of the variational principle, read

$$\partial_b f^{ab} = j^a, \qquad (18)$$

($a$, $b$ = 1... 5), with an additional fifth component of the conserved current $r = |y_t(x)|^2$. The underlying symmetry of the theory can be $O(3,2)$ or $O(4,1)$ "depending on the choice of metric for the raising and lowering of the fifth ($t$) index [35]". For Minkowski-space components the equation (18) is reduced to $\partial_n f^{mn} + \partial_t f^{m5} = j^m$. The Maxwell theory is recovered after integrating over $t$ from $-\infty$ to $\infty$, with appropriate asymptotic conditions. The formalism has been applied mainly in the study of the many-body problem and in the measurement theory, namely, bound states (the hydrogen atom), the scattering problem, the calculation of the selection rules and amplitudes for radiative decays, a covariant Zeeman effect, the Landau-Peierls inequality. Two crucial experiments which may check validity and may distinguish the theory from ordinary approaches have also been proposed [p.15] [37].

Furthermore, one should mention that in the framework of the special relativity version of the Feynman-Dyson proof of the Maxwell's equations [39] S. Tanimura came to rather unexpected conclusions [40] which are related with the formulation defended by L. Horwitz. Trying to prove the Maxwell's formalism S. Tanimura arrived at the conclusion about a theoretical possibility of its generalization. According to his consideration the 4-force acting on the particle in the electromagnetic field must be expressed in terms of

$$F^m(x, \dot{x}) = G^m(x) + \langle F^m{}_n(x)\dot{x}^n \rangle, \qquad (19)$$

where the symbol $<...>$ refers to the Weyl-ordering prescription. The fields $G^m(x)$, $F^m{}_n(x)$ satisfy§

$$\partial_m G_n - \partial_n G_m = 0, \qquad (20a)$$

$$\partial_m F_{nr} + \partial_n F_{rm} + \partial_r F_{mn} = 0. \qquad (20b)$$

This implies that apart from the 4-vector potential $F_{mn} = \partial_m A_n - \partial_n A_m$ there exists a scalar field $f(x)$ such that $G_m = \partial_m f$. One may try to compare this result with the fact of existence of additional scalar field components in the Majorana-Oppenheimer formulation of electrodynamics and with the Stueckelberg-Horwitz theory. The latter has been done by Prof. Horwitz himself [35c] by the identification $F_{m5} = -F_{5m} = G_m$ and the explicit demonstration that for the off-shell theory the Tanimura's equations reduce to

$$\partial_m F_{nr} + \partial_n F_{rs} + \partial_r F_{mn} = 0, \qquad (21a)$$

$$\partial_m G_n - \partial_n G_m + \frac{\partial F_{mn}}{\partial t} = 0, \qquad (21b)$$

$$m\ddot{x}^m = G^m(t, x) + F^{mn}(t, x)\dot{x}_n. \qquad (21c)$$

Finally, among theories with additional parameters one should mention the quantum field model built in the de Sitter momentum space $p_5^2 - p_4^2 - p_3^2 - p_2^2 - p_1^2 = M^2$, ref. [41]. The parameter $M$ is considered as a new physical constant, the fundamental mass. In a configurational space defined on the basis of the Shapiro transformations the equations become the finite-difference equations thus leading to the lattice structure of the space. In the low-energy limit ($M \to \infty$) the theory is equivalent to the standard one.

---

§  One may wish to repeat the Tanimura proof for dual fields and obtain some additional equations.

*Action-at-a-distance.* In the paper [42] A. E. Chubykalo and R. Smirnov-Rueda argued on the basis of the analysis of the Cauchy problem of the D'Alembert and Poisson equations that one should revive the concept of the *instantaneous* action-at-a-distance in classical electrodynamics. The essential feature of the formalism is in introduction of two types of field functions, with the explicit and implicit dependencies on time. The energy of longitudinal modes in this formulation cannot be stored locally in the space, the spread velocity may be whatever and so, they believe, that one has also $E = 0$. The new convection displacement current was proposed in [3] on the basis of the development of this wisdom. It has a form $j_{disp} = -\frac{1}{4p}(\mathbf{v} \cdot \nabla)\mathbf{E}$. This is a resurrection of the Hertz' ideas (later these ideas have been defended by T. E. Phipps, Jr.) to replace the partial derivative by the total derivative in the Maxwell's equations. In my opinion, one can also reveal some connections with the Majorana-Oppenheimer formulation following to the analysis of ref. [p.728] [18].

F. Belinfante [43a] appears to come even earlier to the Sachs' idea about the "physical vacuum" as pairs of some particles from a very different viewpoint. In his formulation of the quantum-electrodynamic perturbation theory zero-order approximation is determined in which scalar and longitudinal photons are present in pairs. He also considered [43b] the Coulomb problem in the frameworks of the quantum electrodynamics and proved that the signal can be transmitted with the velocity greater than *c*. So, this old work appears to be in accordance with recent experimental data (particularly, with the claims of G. Nimtz *et al.* [44] about a wave packet propagating faster than *c* through a barrier, which was used "to transmit Mozart's Symphony No. 40 through a tunnel of 114 mm length at a speed of 4.7*c*"). As indicated by E. Recami in a private communication the $E = 0$ solutions can be put in correspondence to a tachyon of the infinite velocity.

*Evans-Vigier* $\mathbf{B}^{(3)}$ *field.* In a recent series of remarkable papers (in *FPL, FP, Physica A* and *B, Nuovo Cimento B*) and books M. Evans and J.-P. Vigier have indicated the possibility of consideration of the longitudinal $\mathbf{B}^{(3)}$ field for describing many electromagnetic phenomena and in cosmological models as well [12]. It is connected with transverse modes

$$\mathbf{B}^{(1)} = \frac{B^{(0)}}{\sqrt{2}}(i\mathbf{i} + \mathbf{j})e^{if}, \tag{22a}$$

$$\mathbf{B}^{(2)} = \frac{B^{(0)}}{\sqrt{2}}(-i\mathbf{i} + \mathbf{j})e^{-if}, \tag{22b}$$

$f = w\,t - \mathbf{k} \cdot \mathbf{r}$, by means of the cyclic relations

$$\mathbf{B}^{(1)} \times \mathbf{B}^{(2)} = iB^{(0)}\mathbf{B}^{(3)*}, \tag{23a}$$

$$\mathbf{B}^{(2)} \times \mathbf{B}^{(3)} = iB^{(0)}\mathbf{B}^{(1)*}, \tag{23b}$$

$$\mathbf{B}^{(3)} \times \mathbf{B}^{(1)} = iB^{(0)}\mathbf{B}^{(2)*}. \tag{23c}$$

The indices (1), (2), (3) denote vectors connected by the relations of the circular basis and, thus, the longitudinal field $\mathbf{B}^{(3)}$ presents itself a third component of the 3-vector in some isovector space. "The conventional *O*(2) gauge geometry is replaced by a non-Abelian *O*(3) gauge geometry and the Maxwell equations are thereby generalized" in this approach. Furthermore, some success in the problem of the unification of gravitation and electromagnetism has been achieved in recent papers by M. Evans [45]. It has been pointed out by several authors, *e.g.*,[13,46] that this field is the simplest and most natural (classical) representation of a particles spin, the additional phase-free discrete variable discussed by Wigner [9]. The consideration by Y. S. Kim *et al.*, see ref. [47a,formula (14)], ensures that the problem of physical significance of the Evans-Vigier-type longitudinal modes is related with the problem of the normalization and of existence of the mass of a particle transformed on the $(1,0) \oplus (0,1)$ representation of the Poincaré group. Considering explicit forms of the $(1,0) \oplus (0,1)$ "bispinors" in the light-front formulation [49] of the quantum field theory of this representation (the Weinberg-Soper formalism) D. V. Ahluwalia and M. Sawicki [15a] showed that in the massless limit one has only two non-vanishing Dirac-like solutions. The "bispinor" corresponding to the longitudinal solution is directly proportional to the mass of the particle. So, the massless limit of this theory, the relevance of the *E*(2) group to describing physical phenomena and the problem of what is mass deserve further research.

The idea of longitudinal modes related with the electromagnetic field is not so new as it appears at the present time. E. T. Whittaker in the beginning of the century [48] considered the general solution of the D'Alembert wave equation and concluded that "the functions which define the resulting electrodynamic field … can be expressed in terms of the derivatives of *two scalar potential functions*". The direction of corresponding vectors may be chosen in such a way that they are aligned themselves. The physically observable fields are then

$$\mathbf{d} = \text{curl curl } \mathbf{f} + \text{curl } \frac{1}{c} \dot{\mathbf{g}}, \tag{24a}$$

$$\mathbf{h} = \text{curl } \frac{1}{c} \dot{\mathbf{f}} - \text{curl curl } \mathbf{g}, \tag{24b}$$

where **d** and **h** are the electric and magnetic vectors. The field created by arbitrary moving electrons also can be expressed in the terms of **f** and **g**. In modern language, these "longitudinal" functions **f** and **g** (with magnitudes $|\mathbf{f}| = F$, $|\mathbf{g}| = G$) may be related to the Hertz potentials $\mathsf{H}^{mn}$,**

$$F^{mn} = \partial^m \partial_l \mathsf{H}^{ln} - \partial^n \partial_l \mathsf{H}^{lm}. \tag{25}$$

Reducing the Whittaker's general solution to the plane wave (which are in overall use) is straightforward from his formulation.

*Antisymmetric tensor fields*. To the best of my knowledge, modern research into antisymmetric tensor fields in the quantum theory began from the paper by V. I. Ogievetskii and I. V. Polubarinov [50]. They claimed that the antisymmetric tensor field (*notoph* in the terminology used, which I find quite suitable) can be "longitudinal" in the quantum theory, owing to the new gauge invariance

$$F_{mn} \to F_{mn} + \partial_m \Lambda_n - \partial_n \Lambda_m \qquad (26)$$

and applications of the supplementary conditions. The result by Ogievetskii and Polubarinov has been repeated by K. Hayashi [51], M. Kalb and P. Ramond [52] and T. E. Clark *et al.* [53]. The Lagrangian ($F_k = ie_{kjmn} F_{jm,n}$)

$$\mathsf{L}^H = \frac{1}{8} F_k F_k = -\frac{1}{4} (\partial_m F_{na})(\partial_m F_{na}) + \frac{1}{2} (\partial_m F_{na})(\partial_n F_{ma}) \qquad (27)$$

after the application of the Fermi method *mutatis mutandis* (comparing with the case of the 4-vector potential field) yields the spin dynamical invariant to be equal to zero. While several authors insisted on the "transversality" of the antisymmetric tensor field and the necessity of gauge-independent consideration [54-56] perpetually this interpretation ('longitudinality') has become wide-accepted. In refs. [57,58] an antisymmetric tensor *matter* field was studied and it appears to be also "longitudinal", but to have two degrees of freedom. Unfortunately, the authors of the cited work regarded only a massless *real* field and did not take into account the physical reality of the dual field corresponding to an antiparticle. But, what is important, L. Avdeev and M. Chizhov noted [58] that in such a framework there exist $d$-type transverse solutions, which cannot be interpreted as relativistic particles.

---

** Compare this formula with the dynamical equations of the antisymmetric tensor field, *e.g.*, ref.[46,60]. It induces speculations about possible significance of the normalization of the corresponding functions of the momentum representation.

If the antisymmetric tensor field would be pure longitudinal, it appears failure to understand, why in the classical electromagnetism we are convinced that an antisymmetric tensor field is a transverse field. This induces speculations about the incorrectness of the Correspondence Principle. Moreover, this result contradicts with the Weinberg theorem $B - A = 1$, ref. [19b]. This situation has been later analyzed in refs. [59,13,60,46,61] and it was found that indeed the "longitudinal nature" of antisymmetric tensor fields is connected with the application of the generalized Lorentz condition to the quantum states: $\partial_m F^{mn} |\Psi\rangle = 0$. Such a procedure leads also (like in the case of the treatment of the 4-vector potential field without proper regarding the phase field) to the problem of the indefinite metric which was noted by Gupta and Bleuler. So, it is already obviously from methodological viewpoints that the grounds for regarding only particular cases can be doubted by the Lorentz symmetry principles. Ignoring the phase field of Dirac-Fock-Podol'sky-Staruszkiewicz or ignoring $c$ functions [62] related with the 4-current and, hence, with the possible non-zero vacuum value of $\partial_m F^{mn}$ can put obstacles on the way of creation of the unified field theory and embarrass understanding the physical content dictated by the Relativity Theory. This is my opinion.

*The Weinberg formalism.* In the beginning of the sixties the $2(2j+1)$- component approach has been proposed in order to construct a Lorentz-invariant interaction S-matrix from the first principles [63,64,19,65-68]. The authors had thus some hopes on adequate perturbation calculus for processes including higher-spin particles which appeared in the disposition of physicists in that time. The field theory in that time was in some troubles.

The Weinberg *anzatzen* for the $(j,0) \oplus (0,j)$ field theory are simple and obvious [19a,p.B1318]: a) relativistic invariance

$$U[\Lambda, a]\boldsymbol{y}_n(x)U^{-1}[\Lambda, a] = \sum_m D_{nm}[\Lambda^{-1}]\boldsymbol{y}_m(\Lambda x + a), \tag{28}$$

where $D_{mn}[\Lambda]$ is the corresponding representation of $\Lambda$. b) causality

$$[\boldsymbol{y}_n(x), \boldsymbol{y}_m(y)]_\pm = 0 \tag{29}$$

for $(x-y)$ spacelike, which garantees the commutator of the Hamiltonian density $[\mathsf{H}(x), \mathsf{H}(y)] = 0$, provided that $\mathsf{H}(x)$ contains an even number of fermion field factors. The interaction Hamiltonian $\mathsf{H}(x)$ is constructed out of the creation and annihilation operators for the free particles described by some $H_o$, the free-particle part of the Hamiltonian. The $(j,0) \oplus (0,j)$ field

$$\boldsymbol{y}(x) = \begin{pmatrix} \boldsymbol{j}(x) \\ \boldsymbol{c}(x) \end{pmatrix} \tag{30}$$

transforms according to (28), where

$$D^{(j)}[\Lambda] = \begin{pmatrix} D^{(j)}[\Lambda] & 0 \\ 0 & \overline{D}^{(j)}[\Lambda] \end{pmatrix}, \quad D^{(j)}[\Lambda] = \overline{D}^{(j)}[\Lambda^{-1}]^\dagger, \quad D^{(j)}[\Lambda]^\dagger = \boldsymbol{b}D^{(j)}[\Lambda^{-1}]\boldsymbol{b}, \tag{31}$$

with

$$\boldsymbol{b} = \begin{pmatrix} 0 & 1 \\ 1 & 0 \end{pmatrix}, \tag{32}$$

and, hence, for pure Lorentz transformations (boosts)

$$D^{(j)}[L(\mathbf{p})] = \exp(-\hat{\mathbf{p}} \cdot \mathbf{J}^{(j)}q), \tag{33a}$$

$$\overline{D}^{(j)}[L(\mathbf{p})] = \exp(+\hat{\mathbf{p}} \cdot \mathbf{J}^{(j)}q), \tag{33b}$$

with $\sinh q \equiv |\mathbf{p}|/m$. Dynamical equations, which Weinberg proposed, are (Eqs. (7.17) and (7.18) of the first paper [19]):

$$\overline{\Pi}(-i\partial)\boldsymbol{j}(x) = m^{2j}\boldsymbol{c}(x), \tag{34a}$$

$$\Pi(-i\partial)\boldsymbol{c}(x) = m^{2j}\boldsymbol{j}(x). \tag{34b}$$

These are rewritten into the form (Eq. (7.19) of [19a])

$$\left[\boldsymbol{g}^{m_1 m_2 \ldots m_{2j}} \partial_{m_1} \partial_{m_2} \ldots \partial_{m_{2j}} + m^{2j}\right]\boldsymbol{y}(x) = 0, \tag{35}$$

with the Barut-Muzinich-Williams matrices [63]

$$\boldsymbol{g}^{m_1 m_2 \ldots m_{2j}} = -i^{2j} \begin{pmatrix} 0 & t^{m_1 m_2 \ldots m_{2j}} \\ \bar{t}^{m_1 m_2 \ldots m_{2j}} & 0 \end{pmatrix}. \tag{36}$$

The following notation was used

$$\Pi^{(j)}_{s's}(q) \equiv (-1)^{2j} t_{s's}{}^{m_1 m_2 \ldots m_{2j}} q_{m_1} q_{m_2} \ldots q_{m_{2j}}, \tag{37}$$

$$\overline{\Pi}^{(j)}_{s's}(q) \equiv (-1)^{2j} \bar{t}_{s's}{}^{m_1 m_2 \ldots m_{2j}} q_{m_1} q_{m_2} \ldots q_{m_{2j}}, \quad \overline{\Pi}^{(j)*}(q) = C\Pi^{(j)}C^{-1}, \tag{38}$$

with *C* being the matrix of the charge conjugation in the $2j + 1$- dimension representation (cf. [14]). The tensor *t* is defined in a following manner:

- $t_{s's}{}^{m_1 m_2 \ldots m_{2j}}$ is a $2j + 1$ matrix with $s, s' = +j, j-1, \ldots, -j; m_1, m_2 \ldots m_{2j} = 0, 1, 2, 3$;

- *t* is symmetric in all *m*'s;

- *t* is traceless in all *m*'s, i.e., $g_{m_1 m_2} t_{s's}{}^{m_1 m_2 \ldots m_{2j}}$ and with all permutations of upper indices;

- *t* is a tensor under Lorentz transformations,

$$D^{(j)}[\Lambda] t^{m_1 m_2 \ldots m_{2j}} D^{(j)}[\Lambda]^\dagger = \Lambda^{m_1}_{n_1} \ldots \Lambda^{m_{2j}}_{n_{2j}} t^{n_1 n_2 \ldots n_{2j}}, \tag{39a}$$

$$\overline{D}^{(j)}[\Lambda] \overline{t}^{m_1 m_2 \ldots m_{2j}} \overline{D}^{(j)}[\Lambda]^\dagger = \Lambda^{m_1}_{n_1} \ldots \Lambda^{m_{2j}}_{n_{2j}} \overline{t}^{n_1 n_2 \ldots n_{2j}} \tag{39b}.$$

For instance, in the $j = 1$ case $t^{00} = 1$, $t^{0i} = t^{i0} = J_i$ and $t^{ij} = \{J_i, J_j\} - \delta_{ij}$, with $J_i$ being the $j = 1$ spin matrices and the metric $g_{mn} = \text{diag}(-1,1,1,1)$ being used. Furthermore, for this representation

$$\overline{t}^{m_1 m_2 \ldots m_{2j}} = \pm \overline{t}^{m_1 m_2 \ldots m_{2j}}, \tag{40}$$

the sign being +1 or –1 according to whether the *m*'s contain altogether an even or an odd number of space-like indices.

The Feynman diagram technique has been built and some properties with respect to discrete symmetry operations have been studied. The propagator used in the Feynman diagram technique is found not to be the propagator arising from the Wick theorem because of extra terms proportional to equal-time $\delta$-functions and their derivatives appearing if one uses the time-ordering product of field operators $\langle T\{\psi_a(x)\overline{\psi}_b(y)\}\rangle_0$. The covariant propagator is defined

$$S_{ab}(x-y) = (2\pi)^{-3} i m^{-2j} M_{ab}(-i\partial) \int \frac{d^3\mathbf{p}}{2w(\mathbf{p})} \left[\begin{array}{c} \theta(x-y)\exp\{ip\cdot(x-y)\} + \\ +\theta(y-x)\exp\{ip\cdot(y-x)\} \end{array}\right], \tag{41}$$

$$= -m^{-2j} M_{ab}(-i\partial) \Delta^C(x-y)$$

where

$$M(p) = \begin{pmatrix} m^{2j} & \Pi(p) \\ \overline{\Pi}(p) & m^{2j} \end{pmatrix}, \tag{42}$$

and $\Delta^C(x)$ is the covariant $j = 0$ propagator.

For massless particles the Weinberg theorem about connections between the helicity of a particle and the representation of the group (*A,B*) which the corresponding field transforms on has been proved. It says:

> "A massless particle operator $a(\mathbf{p}, \lambda)$ of helicity $\lambda$ can only be used to construct fields which transform according to representations (A, B), such that $B - A = \lambda$. For instance, a left-circularly polarized photon with $\lambda = -1$ can be associated with (1,0), (3/2, 1/2), (2,1)... fields, but not with the vector potential, (½,½)... [It is not the case of a massive particle.] A field can be constructed out of $2j + 1$ operators $a(\mathbf{p}, \sigma)$ for any representation (A, B) that "contains" j, such that $j = A + B$, or $A + B - 1$... or

*|A −B|, [e.g., a j = 1 particle massive] field could be a four-vector (½,½)…[i.e., built out of the vector potential ]."*

In subsequent papers Weinberg showed that it is possible to construct fields transformed on other representations of the Lorentz group but unlikely can be considered as fundamental ones. The prescription for constructing the fields have been given in ref. [19c,p.1895].

*"Any irreducible field $y^{(A,B)}$ for a particle of spin j may be constructed by applying a suitable differential operator of order 2B to the field $y^{(j,0)}$, provided that A, B, and j satisfy the triangle inequality |A − B| = j = A + B."*

For example, from the self-dual antisymmetric tensor $F^{\mu\nu}$ the (½,½) field $\partial_m F^{mn}$, the (0,1) field $e_{mnl\,t} \partial^l \partial_s F^{rs}$ have been constructed. Moreover, various invariant-type interactions have been tabulated [19b,p.B890] and [19c,Section III]. While one can also use fields from different representations of the Lorentz group to obtain some physical predictions, in my opinion, such a wisdom could lead us to certain mathematical inconsistencies (like the indefinite metric problem and the subtraction of infinities [4]). The applicability of the procedure mentioned above to massless states should still be analyzed in detail.

Finally, in another paper [66c] Weinberg wrote:

*"Tensor fields cannot by themselves be used to construct the interaction H'(t)… The potentials are not tensor fields… It is for this reason that some field theorists have been led to introduce fictitious photons and gravitons of helicity other than ±j, as well as the indefinite metric that must accompany them. Preferring to avoid such unphysical monstrosities, we must ask now what sort of coupling we can give our nontensor potentials without losing the Lorentz invariance of the S matrix?… Those in which the potential is coupled to a conserved current."*

Thus, he tried to provide some basis to the gauge models from the Lorentz invariance. In the recent book [69] he slightly changed his views:

*"Interactions in such a theory [constructed from $f^{mn}$ and its derivatives] will have a rapid fall-off at large distances, faster than the usual inverse-square law. This is perfectly possible, but… theories that use vector fields for massless spin one particles represent a more general class of theories…"*

My opinion is: all reliable theories must have well-defined massless limit and be in accordance with the Weinberg theorem. While many recipes were developed to handle with interactions mediated by virtual particles described by the 4-vector potential, the questions, which theories "are actually realized in nature" and which is the more general theory, remain to be opened.

We have reached this conclusion on the basis of our development of the Weinberg theory [19,66] and its reformulation by A. Sankaranarayanan [70];[††] of the Majorana con-cept of the neutrality [71] and its reformulation by J. A. McLennan and K. M. Case [72]; and after reading the important work of B. Nigam and L. Foldy [73] and useful suggestions of the referee [11] of the work [14b]. Here we are not going to discuss our recent work in detail and only list some important results:

- It was proposed another equation in the $(1,0) \oplus (0,1)$ representation space [70]:

$$\left[ g_{mn} p_m p_n + \frac{i(\partial/\partial t)}{E} m^2 \right] \psi = 0. \tag{43}$$

---

[††] Unfortunately, the author of the cited work did not realized in 1965 himself that his equation describes particles with different physical properties compared with the initial Weinberg formulation.

  In such a framework a boson and its antiboson have *opposite* intrinsic parities [14]. The conclusion was also reached in the Fock space. The essential feature in deriving the equation (ref. [70]) in ref. [14] was the Ryder-Burgard relation in the form $f_R = \pm f_L$. The presented theory [14b] is the first explicit example of the theory of the Bargmann-Wightman-Wigner type [9b].

- The concept of the *complex* $(1,0) \oplus (0,1)$ fields as parts of the degenerate doublet was proposed (ref. [60] and private communication from D. V. Ahluwalia). The representation is spanned by the two six-component functions in the coordinate space (*e.g.*, $\psi_1$ and $g\psi_1$, or $F^{mn}$ and $\widetilde{F}^{mn}$). The mapping between the antisymmetric tensor and Weinberg formulations has been found. Properties of the field functions with respect to $P \equiv g_4$ operation have been studied. The dynamical invariants for the Weinberg field [60] and for the antisymmetric tensor field [46] have been obtained.

- The boson-boson interaction amplitude appears to be very similar [74] to the fermion-fermion amplitude in the second order of the Feynman perturbation theory if one works in the Lobachevsky momentum space. The only difference is that the denominator in the former has to be changed: $1/\vec{\Delta}^2 \to 1/2m(\Delta_o - m)$. The spin structure of the numerator remains to be unchanged. $(\Delta_o, \vec{\Delta})$ is the 4-vector of the momentum transfer in the Lobachevsky space, see, *e.g.*, ref. [75].

- The Majorana-Oppenheimer formulation [see equations (5a,b)] has been generalized to the massive field case [76] by using some ideas of the paper [77].

- The relativistic covariance of the **B**- cyclic relations has been proven [78].

- In the $(½,0) \oplus (0,½)$ representation space the self/anti-self charge conjugate spinors $\lambda^{S,A}(p^m)$ [and related to them $\rho^{S,A}(p^m)$ spinors] have been introduced [16]. One can note interesting features of these spinors: they are not eigenspinors of the parity operator; they are not eigenspinors of the helicity operator $h$ of the $(½,0) \oplus (0,½)$ representation (but, the new operator $-gh$ of the *chiral helicity* was introduced); for massless particles $\lambda^{S,A}_\uparrow$ identically vanish; they form *bi-orthonormal* set in the mathematical sense (see formula (41) in [16]). In the $(1,0) \oplus (0,1)$ representation it is impossible to construct self/anti-self charge conjugate objects in the similar way. But, the eigenvectors of the $\Gamma_5 S^c_{[1]}$ operator have been introduced there [16].

- The Majorana representation (MR) and the corresponding unitary matrix $y^M = Uy^W$ of the transfer to the MR have been defined [79]. In this representation $l^S$ and $r^A$ keep to be pure *real* and $l^A$ and $r^S$ keep to be pure *imaginary* for both spin-½ and spin-1 case (*cf.* [71]).

- On the basis of the generalization of the Ryder-Burgard relation (see the formula (11) in [17b]) the dynamical equations in the covariant form have been derived in both the (½,0) ⊕ (0,½) and (1,0) ⊕ (0,1) representation spaces [17]. The explicit form of these "MAD" equations in the $j = $ ½ case is

$$ig^m\partial_m l^S(x) - mr^A(x) = 0, \quad ig^m\partial_m r^A(x) - ml^S(x) = 0, \tag{44a}$$

$$ig^m\partial_m l^A(x) + mr^S(x) = 0, \quad ig^m\partial_m r^S(x) + ml^A(x) = 0. \tag{44b}$$

A fermion and its antifermion appear to be able to carry the *same* intrinsic parities [80] in the framework of the similar construction in the Fock space. So, the Bargmann-Wightman-Wigner-type quantum field theory [9b] can be realized in the (½,0) ⊕ (0,½) representation space too.

- Gauge transformations for the $l^{S,A}$ and $r^{S,A}$ spinors take the form, ref. [17b]

$$l' \to (\cos a - ig^5 \sin a)l(x), \tag{45a}$$

$$r' \to (\cos a + ig^5 \sin a)r(x). \tag{45b}$$

Thus, we have automatically parity-violating currents.

- It is interesting to note that oscillations such as $\langle l^A_{-h} | l^S_h(t) \rangle \sim \sin^2(Et/\hbar)$ are possible. This induces a lot of speculations on the foundations of quantum mechanics.

- Constructs presented in refs. [81,82] seem to have similar physical content comparing with the Majorana-Ahluwalia construct. The authors of [82] also proposed the *doubling* of the Fock space; investigated field functions which are not the eigenvectors of the parity operator, while are the eigenvectors of the operator of charge conjugation (defined in a different way). They also regarded the *pseudoscalar* charge. At last, they argued that "The usual '*CP*-mirror' symmetry of the weak interaction should *quite generally* be re-interpretable as a pure *P*-mirror one. The result is that *now the P-mirror image of the actual process* $n \to p + e + \bar{n}$ *should just be identified with the actual antiprocess* $\bar{n} \to \bar{p} + \bar{e} + n$.

- Finally, in the papers [83] it was shown that the solutions of the Maxwell equations and the Klein-Gordon equation (and presumably other relativistic equations) are not necessarily required to be plane waves.

### Acknowledgments

Discussions with Profs. D. V. Ahluwalia, A. E. Chubykalo, M. W. Evans, H. Múnera, I. Nefedov, A.F. Pashkov, S. Roy and M. Sachs were invaluable for writing the paper. I greatly appreciate many private communications and preprints from colleagues over the world.

I am grateful to Zacatecas University for a professorship. This work has been supported in part by the Mexican Sistema Nacional de Investigadores, the Programa de Apoyo a la Carrera Docente and by the CONACyT, México under the research project 0270P-E.